\documentclass[final,3p,times,twocolumn]{elsarticle}
\biboptions{sort&compress}

\usepackage{graphicx}

\newcommand{\tc}{\ensuremath{T_{\it c}}}
\newcommand{\hw}{\ensuremath{\hbar \omega}}
\newcommand{\ibbfcax}{Ba(Fe$_{1-x}$Co$_x$)$_2$As$_2$}
\newcommand{\ibbs}[1]{\mbox{\boldmath \ensuremath{#1}}}

\newcommand{\ibwlo}{\ensuremath{F_{\rm{lo}}}}
\newcommand{\ibwtr}{\ensuremath{F_{\rm{tr}}}}

\journal{Physica C}

\begin{document}

\begin{frontmatter}

\title{Anisotropic inplane spin correlation in the parent and Co-doped BaFe$_2$As$_2$: a neutron scattering study}

\author[issp,jst]{S.~Ibuka\corref{cor1}\fnref{pakek}}
\ead{ibuka@post.j-parc.jp}
\author[issp,jst]{Y.~Nambu\fnref{paimram}}
\author[issp,jst]{T.~Yamazaki\fnref{patus}}
\author[ornl]{M.~D.~Lumsden}
\author[issp,jst]{T.~J.~Sato\fnref{paimram}}

\address[issp]{Neutron Science Laboratory, Institute of Solid State Physics, University of Tokyo, Tokai, Ibaraki 319-1106, Japan}
\address[jst]{TRIP, JST, Chiyoda, Tokyo 102-0075, Japan}
\address[ornl]{Oak Ridge National Laboratory, Oak Ridge, TN 37831, USA}

\cortext[cor1]{Corresponding author}
\fntext[pakek]{Present address: High Energy Accelerator Research Organization, Tokai, Ibaraki 319-1106, Japan.}
\fntext[paimram]{Present address: Institute of Multidisciplinary Research for Advanced Materials, Tohoku University, Katahira, Sendai 980-8577, Japan.}
\fntext[patus]{Present address: Faculty of Science and Technology, Tokyo University of Science, Noda, Chiba 278-8510, Japan.}

\begin{abstract}
Antiferromagnetic spin fluctuations were investigated in the normal states of the parent ($x = 0$), under-doped ($x = 0.04$) and optimally-doped ($x = 0.06$) {\ibbfcax} single crystals using inelastic neutron scattering technique. 
For all the doping levels, quasi-two-dimensional antiferromagnetic fluctuations were observed as a broad peak localized at $\ibbs{Q} = (1/2, 1/2, l)$. 
At lower energies, the peak shows an apparent anisotropy in the $hk0$ plane; longitudinal peak widths are considerably smaller than transverse widths.
The anisotropy is larger for the higher doping level. 
These results are consistent with the random phase approximation (RPA) calculations taking account of the orbital character of the electronic bands,
confirming that the anisotropic nature of the spin fluctuations in the normal states is mostly dominated by the nesting of Fermi surfaces. 
On the other hand, the quasi-two-dimensional spin correlations grow much rapidly for decreasing temperature in the $x = 0$ parent compound, compared to that expected for nearly antiferromagnetic metals. This may be another sign of the unconventional nature of the antiferromagnetic transition in BaFe$_2$As$_2$.
\end{abstract}

%PACS

%%\pacs{74.70.Xa, 74.72.Ek, 75.40.Gb}

\end{frontmatter}

\section{Introduction}
For iron-based superconductors~\cite{Kam08}, the conventional theory of phonon-mediated superconductivity has difficulty in explaining the high superconducting transition temperatures~\cite{Boer08, Maz08, Boer10}.
Accordingly, various other candidates for the superconducting pairing mechanism have been proposed to date, such as spin-fluctuation mediated $s_{+-}$ model~\cite{Maz08, Emer64, Berk66, Kur08}, as well as orbital-fluctuation mediated $s_{++}$ model~\cite{Sta08, Kon10, Yan10, Yin10}.
To determine which pairing mechanism is indeed appropriate, it is crucial to know the details of the spin and orbital fluctuations in the normal paramagnetic state.
Since direct observation of the orbital fluctuations is difficult, experimental efforts have been focused on observation of the spin fluctuations using neutron inelastic scattering technique.
Among a number of Fe-based superconductor compounds, 
$A$Fe$_2$As$_2$ ($A =$ Ca, Sr, Ba and K) 122-type compounds have been most intensively studied due to the availability of large single crystals with various doping levels.
Both in the parent and doped compounds, rod-like low energy spin excitation with weak spin correlation along $l$ was observed around the zone boundary $\ibbs{Q} = (1/2, 1/2, l)$ in the tetragonal paramagnetic state, for example, in {\ibbfcax} ($0 \leq x \leq 0.08$)~\cite{Pra09, Chr09, Lum09, Mat09, Ino10, Pra10, Mat10}.
The $\ibbs{Q}$ vector
connects hole Fermi surface sheets at the antiferromagnetic zone centre to electron sheets at the zone corner, 
and satisfies the nesting condition. 
Moreover, in the heavily-overdoped {\ibbfcax} ($x = 0.24$), the inelastic excitation disappears~\cite{Mat10}\, and the angle-resolved photoemission spectroscopy observed that the hole pockets disappear in 0.15 $< x <$ 0.3~\cite{Bro09, Sud10}.
These results suggest that the low-energy spin excitation originates from the Fermi surface nesting between the hole and electron sheets.

Recent studies~\cite{Dia10, Les10, Li10, Par10, Lee11, Harr11, Zhan11, Ewin11, Luo12, Tuck12, Liu12, Harr12, Luo13, Wang13} have detected clear inplane anisotropy in the spin correlation lengths.
At low energies, the rod-like peak appears with elliptical cross section in the two-dimensional $hk0$ plane, having longer axis pointing to the transverse direction in the parent and electron doped compounds, whereas pointing to the longitudinal direction in the hole doped compounds.
At high energies (${\hw} >$ 80 meV), the elliptical peak enlarges and splits, with no clear change on entering in the orthorhombic phase.
As Park {\it et al} explained~\cite{Par10}, the anisotropy preserves C$_4$ symmetry with the symmetry axis (0, 0, $l$), and is different from rotational symmetry breaking.
They suggest~\cite{Par10, Gra10} that such inplane anisotropy, at least in the low-energy range, can be consistently reproduced by a simple random phase approximation (RPA) calculation taking account of orbital characters.
It has been an issue if such a anisotropic spin correlations may be naturally attributed to the multiband nature of the Ba-122 compounds, or much intriguing idea has to be introduced, such as the frustrated $J_1-J_2$ model~\cite{Dia10, Li10}, quasi propagating mode with different velocity~\cite{Li10}, and interplay between anisotropies of the correlation length and of Landau damping~\cite{Tuck12, ParkH11}. 
Above controversy may be due to the lack of consistent dataset in one compound family; Park {\it et al}~\cite{Par10} compared the low energy anisotropy in {\ibbfcax} ($x = 0.075$) to that in CaFe$_2$As$_2$~\cite{Dia10}, whereas another comparison was made with the hole doped KFe$_2$As$_2$~\cite{Lee11}.
Hence, it is obvious that direct comparison between the parent compound and electron doped compound, such as BaFe$_2$As$_2$ and Ba(Fe,Co)$_2$As$_2$, under the same condition is essential. 
Luo {\it et al} elaborately studied the anisotropy both in the antiferromagnetic and paramagnetic phase in Ba(Fe$_{1-x}$Ni$_x$)$_2$As$_2$ crystals ($0.015 \leq x \leq 0.09$)~\cite{Luo12}, however it is hard to see the doping dependence of the anisotropy in the paramagnetic phase under the same energy and temperature. 
Therefore, in this work, we performed electron-doping dependence study of the inplane anisotropy of low-energy spin fluctuations in \ibbfcax\ crystals ($x = 0$, 0.04 and 0.06) by inelastic neutron scattering, focusing on the paramagnetic phase.
We observed clear anisotropic inplane spin correlations for all the doping levels. 
The anisotropy in BaFe$_2$As$_2$ is smaller than the electron doped Ba(Fe, Co)$_2$As$_2$. 
This result is consistent with the Fermi surface nesting picture and indicates that the anisotropic nature of the spin fluctuations in the low energy regime are dominated by Fermi surface nesting. 
Concerning the temperature dependence of the peak width, the doped compounds show consistent behaviour expected for nearly antiferromagnetic metals, whereas the peak in the parent compound sharpens much pronouncedly.
This suggests that the quasi-two-dimensional spin correlations grow much rapidly for decreasing temperature in the $x = 0$ parent compound.
This may be another sign of the unconventional nature of the antiferromagnetic transition in BaFe$_2$As$_2$.

\section{Experimental details}
Single crystals of \ibbfcax\ ($x = 0$, 0.04(\#1), 0.04(\#2), and 0.06) were synthesized in a FeAs self-flux using the Bridgman method. 
First, FeAs precursor was prepared from 99.9\% Fe and 99.9999\% As powders. 
The starting elements were mixed, put into an alumina crucible, and sealed in a quartz tube under an argon gas atmosphere. 
Then the starting elements were slowly heated up to 1073~K, and kept there for 5~days.

Next, 99.9\% Ba chips, 99.9\% Co powder, 99.9999\% As powder, and the prepared FeAs powder, weighed with a molar ratio of Ba:(Fe, Co):As as 10:45:45, were mixed, put in a carbon crucible, sealed in a tantalum crucible, and were further sealed in a quartz tube.
All the procedures were performed in an argon-filled glove box with O$_2$ concentration being about 1~ppm to avoid oxidation.
The sealed starting elements were then set in the vertical Bridgman furnace to obtain large single crystals; details of the Bridgman technique used in this study are given in~\cite{Mor09}. 
We performed the Bridgman growth four times for different Co compositions, and each batch obtained was found to contain several small pieces of single crystals. The mass of the grown pieces of single crystals was between 0.3 and 1.2~grams. 

\begin{figure}
	\begin{center}
        \includegraphics[width=0.9\hsize]{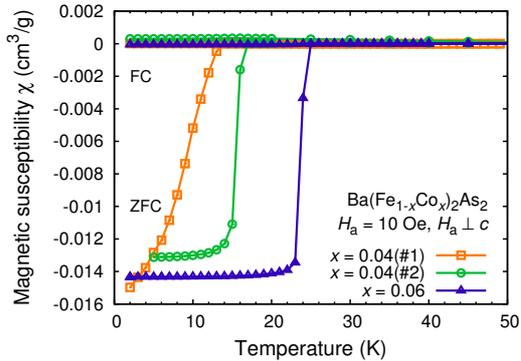}
        \caption{\label{fig1}(Colour online) Temperature dependence of zero-field-cooled (ZFC) and field-cooled (FC) dc magnetic susceptibility of \ibbfcax. The (orange) squares stand for $x = 0.04$(\#1), (green) circles for 0.04(\#2) and (blue) triangles for 0.06. The solid lines are guide to the eyes.}
    \end{center}
\end{figure}

Co compositions of the obtained crystals were determined by energy dispersive X-ray analysis using a scanning electron microscopy JEOM JSM-5600 and Oxford Link ISIS. 
The resulting sample compositions are $x_{\rm EDX} = 0.04$(1), 0.04(1) and 0.061(4) for $x = 0.04$(\#1), 0.04(\#2) and 0.06 samples. The onset temperatures of the superconducting transition were confirmed by dc magnetic susceptibility measurements using a superconducting quantum interference device magnetometer Quantum Design MPMS-XL in an applied magnetic field of 10~Oe perpendicular to the {\it c} axis. 
Figure~\ref{fig1} shows the obtained magnetic susceptibility in the low temperature region.
As seen from the susceptibility data, the superconducting transition temperatures are 13, 16 and 24~K, for the doped three samples $x = 0.04$(\#1), 0.04(\#2) and 0.06, respectively.
Antiferromagnetic transition temperatures were also found to be $T_{\rm{AF}} \sim$ 140, 70 and 70~K for $x = 0$, 0.04(\#1) and 0.04(\#2) crystals, respectively. 
Co compositions determined by {\tc} with the help of the previous report~\cite{Chu09} are consistent with those determined by energy dispersive X-ray analysis; $x_{\tc} \sim 0.040$--0.035, 0.045 and 0.06 for $x = 0.04$(\#1), 0.04(\#2) and 0.06, respectively.
In this study, we regard the two $x = 0.04$ samples (\#1 and \#2) as one composition. Although slight difference in the compositions for the two samples makes considerable change in the superconducting transition temperatures, we believe that such slight composition difference does not give rise to any significant difference in the inelastic response in the paramagnetic phase. This treatment will be accepted by the weak dependence of $F$ and $\Gamma$ parameters on the composition, as we see below. 

Using the four crystals, we performed inelastic neutron scattering experiments. 
We used two thermal neutron triple-axes spectrometers, ISSP-GPTAS installed at JRR-3, Tokai, Japan and HB3 installed at HFIR at Oak Ridge National Laboratory, TN, USA. 
The parent compound ($x = 0$) and underdoped compound ($x = 0.04$(\#1)) were measured at GPTAS. The same parent compound, the other underdoped compound ($x = 0.04$(\#2)) and the optimally-doped compound ($x = 0.06$) were measured at HB3.
Pyrolytic graphite 002 reflections were used both for the monochromator and analyzer to select an energy of neutrons.
Final neutron energy was set to ${\it E_f} = 14.7$~meV.
To detect small differences in peak width, collimations of 40'-80'-40'-80' at GPTAS and 48'-80'-40'-90' at HB3 were employed for most of the measurements. 
The selection of the similar collimations at the two spectrometers enables us to compare the ratio of the anisotropy reliably throughout the investigated compositions.
At large energy transfers, such as $\hw = 28$~meV, where signal becomes weaker, horizontal focusing monochromator with 40'-3 blades Radial Collimator (3RC)-80'-80' was employed at GPTAS.
Higher harmonic neutrons were eliminated by using pyrolytic graphite filters.

To obtain sufficient intensity, two or three pieces of single crystals were co-aligned; the total mass of the samples was about 1~g for all the doping levels. 
Mosaic spreads of the co-aligned samples in the scattering plane were within 1.2, 0.5, 0.7 and 0.6 degrees of full width at half maximum for the samples of $x = 0$, 0.04(\#1), 0.04(\#2) and 0.06, respectively. 
The broadening of the peak width along transverse $\ibbs{Q}$ direction due to the sample mosaic was negligible compared with that due to the instrumental resolution. 
The co-aligned crystals were sealed in aluminum cans and then set in closed cycle $^4$He refrigerators.

\section{Results}
\subsection{Doping dependence of anisotropy at ${\hw} = 10$~{\rm meV}}

\begin{figure*}
\begin{center}
  \includegraphics[width=0.98\hsize]{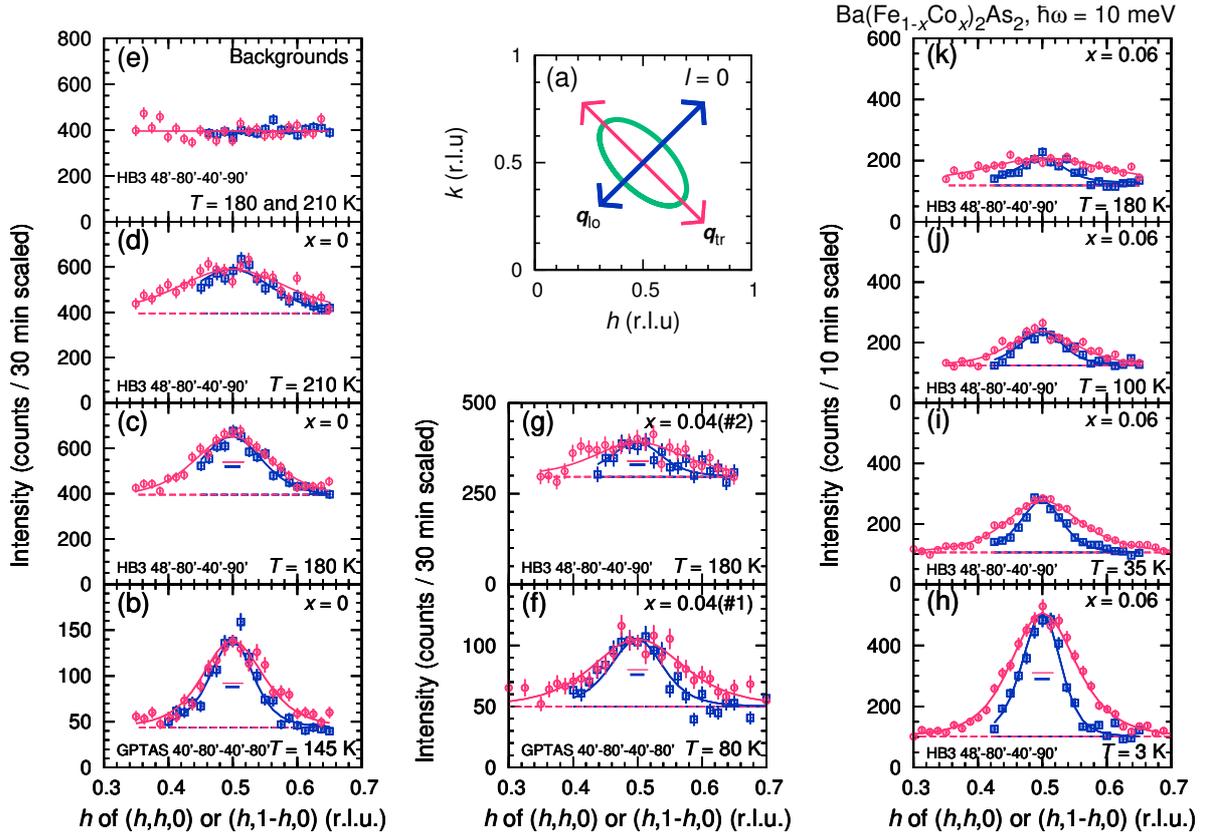}  
  \caption{\label{fig2}(Colour online) Longitudinal and transverse scans around $\ibbs{Q} = (1/2, 1/2, 0)$ in the normal paramagnetic state of {\ibbfcax} at ${\hw} = 10$~meV. (a) Schematic drawing of the directions of the scans. The short thick (blue) arrow indicates the longitudinal direction, and the long thin (red) arrow the transverse direction. (b)--(d) Scans of $x = 0$ at $T = 145$, 180 and 210~K. The (blue) squares stand for the longitudinal scans, and (red) circles the transverse scans. The thick (blue) and thin (red) bars denote the instrumental resolutions along the longitudinal and transverse directions, respectively. The solid lines indicate fits. The dashed lines show backgrounds in the fitting procedure. (e) Backgrounds of $x = 0$ measured at HB3. (f) Scans of $x = 0.04$(\#1) at $T = 80$~K. (g) Scans of $x = 0.04$(\#2) at 180~K. (h)--(k) Scans of $x = 0.06$ at $T = 3$, 35, 100 and 180~K. Data (b)--(d), (f)--(g) and (h)--(k) were measured with a counting time of more than 30, 30 and 10 min, respectively, but are normalized.} 
\end{center}
\end{figure*}
First, doping dependence of anisotropy was investigated by measuring the inelastic scattering peak along the longitudinal direction $(h, h, 0)$ and transverse direction $(h, -h, 0)$ around $\ibbs{Q} = (1/2, 1/2, 0)$ at the low energy transfer ${\hw} = 10$~meV.
Figure~\ref{fig2}(a) shows direction of the scans in the $hk0$ plane.
The results of the constant-energy scans of the parent compound $x = 0$ in the paramagnetic phase at $T = 145$, 180 and 210~K are shown in Figs.~\ref{fig2}(b), (c) and (d), respectively.
Background for HB3 was estimated by performing a similar scan with the crystal angle rotated by 40~degrees, and is shown in figure~\ref{fig2}(e).
For GPTAS, scans with an empty sample-can were performed.
The instrumental resolutions for the longitudinal and transverse scans are indicated by the thick (blue) and thin (red) solid bars in the figures, respectively.
Apparently, the transverse scans show broader peak widths, compared to the longitudinal scans, and hence the antiferromagnetic correlations in the $ab$ plane are anisotropic at ${\hw} = 10$~meV.
It should be noted that the instrumental resolutions are sufficiently narrow, so that the apparent difference in the peak widths cannot be due to the resolution effect.
(This point will be further confirmed by the resolution convoluted fitting later.)

For the underdoped compounds $x = 0.04$(\#1) and 0.04(\#2), the results of constant-energy scans in the paramagnetic phase at $T = 80$ and 180~K are shown in Figs.~\ref{fig2}(f) and (g), respectively. 
By comparing the peak widths at the same temperature 180~K, we found that the longitudinal width in the underdoped compound is mostly the same as that observed in the parent compound, whereas the transverse width becomes significantly wider.
This indicates that the corresponding antiferromagnetic correlations become more anisotropic in the $ab$ plane.

For the optimally doped compound $x = 0.06$, Figs.~\ref{fig2}(h), (i), (j) and (k) show scans in the superconducting phase at $T = 3$~K and in the normal phase at 35, 100 and 180~K, respectively.
Again, the transverse widths are significantly larger than the longitudinal widths, and the width at 180~K in the optimally doped compound exhibits significant broadening compared even to the underdoped one.
When the system is cooled into the superconducting phase at $T = 3$~K, the peak intensity becomes about twice larger than those in the normal state. 
This is due to the enhancement of antiferromagnetic fluctuations in the superconducting phase, reported repeatedly in the Fe-based superconductors~\cite{Lum09, Li10}.

From these results, we can conclude that the peaks are definitely anisotropic in the $ab$ plane for all the doping levels; the widths along the transverse direction are considerably wider than those along the longitudinal direction. The anisotropy seems to be enhanced for the higher doping level.

To estimate the anisotropy further quantitatively, the data were fitted to a model scattering function derived from the generalized susceptibility of nearly antiferromagnetic metals for small $\ibbs{q}$~\cite{Ino10, Dia10, Mor85}:
\begin{eqnarray}
&&I(\ibbs{Q}, \omega) \propto \frac{{\chi}''(\ibbs{q}, \omega)}{1 - \exp \left[- \hw / (\rm{k}_B T) \right] },\\
&&\label{eq:chi} {\chi}''(\ibbs{q}, \omega) =\nonumber\\
&&\frac{\chi_0(T) \Gamma(T) {\hw}}{({\hw})^2 + \Gamma(T)^2 \left( 1 + \ibwlo^2 q_{\rm{lo}}^2+ \ibwtr^2 q_{\rm{tr}}^2 + D^2 q_{\it c}^2 \right)^2},
\end{eqnarray}
where
\begin{eqnarray}
\ibbs{q} &=& \ibbs{Q} - \ibbs{Q}_{\rm{AF}}\\
&=& \left[ 1/\sqrt{2} \left(q_{\rm{lo}} + q_{\rm{tr}}\right), 1/\sqrt{2} \left(q_{\rm{lo}} - q_{\rm{tr}}\right), q_{\it c} \right].
\end{eqnarray}
$\chi_0(T)$ represents the isothermal susceptibility, $\Gamma(T)$ is the isotropic damping constant, and $q_{\rm{lo}}$, $q_{\rm{tr}}$ and $q_{\it c}$ are the norms of the wave vectors away from an antiferromagnetic zone centre $\ibbs{Q}_{\rm{AF}}$ along $(h, h, 0)$, $(h, -h, 0)$ and $(0, 0, l)$, respectively. 
{\ibwlo}, {\ibwtr} and $D$ are the inverse of the peak widths along the three directions, corresponding to the magnetic correlation lengths. 

In the fitting procedure, the model scattering function was convoluted by the instrumental resolution function~\cite{Coo67}, and was used to fit the background-subtracted data.
Temperature dependence of $\Gamma$ was constrained to obey the linear form $\Gamma(T) = \alpha(T + \Theta)$ where $\alpha = 0.14$~meV/K and $\Theta = 30$~K~\cite{Ino10, Mat10}.
$D$ was set to 1.3 r.l.u. and was assumed to be temperature independent~\cite{Mat10}.
The backgrounds were set to $\ibbs{Q}$ and $T$ independent for $x = 0$, whereas for $x = 0.04$(\#1), 0.04(\#2) and 0.06, $\ibbs{Q}$ independent but slightly $T$ dependent.
The solid lines in figure~\ref{fig2} indicate fits to (\ref{eq:chi}), and the dashed lines denote fitted backgrounds.
All peaks were fitted well with the above model function.

\begin{figure}
\begin{center}
  \includegraphics[width=0.9\hsize]{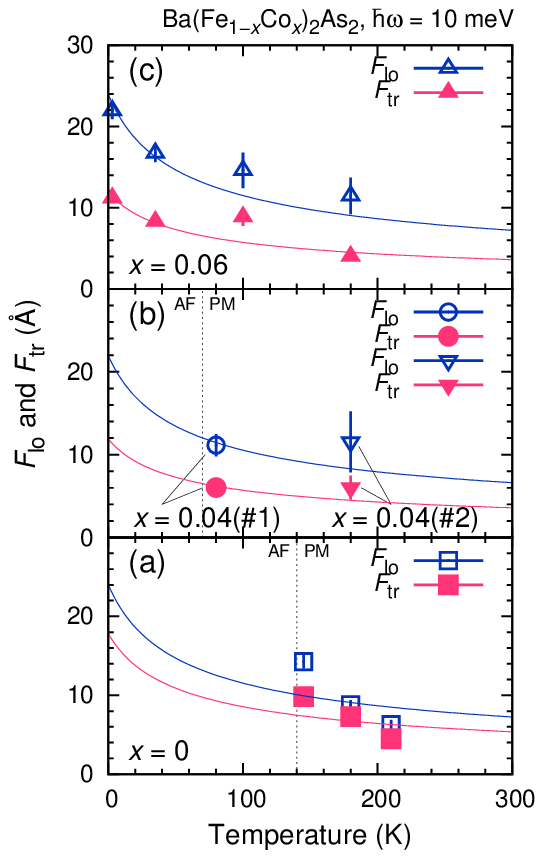}
  \caption{\label{fig3}(Colour online) Temperature dependence of $\ibwlo$ and $\ibwtr$ of $\ibbs{Q} = (1/2, 1/2, 0)$ at ${\hw} = 10$~meV in {\ibbfcax} for (a) $x = 0$, (b) $0.04$(\#1) and 0.04(\#2) and (c) 0.06. The vertical (black) dotted lines in (a) and (b) stand for $T_{\rm AF} = 140$ and 70~K, respectively. The solid lines show fits with $(T+\Theta)^{-1/2}$. 
}
\end{center}
\end{figure}
Temperature dependence of obtained optimum parameters \ibwlo\ and \ibwtr\ is shown in figure~\ref{fig3}(a), (b) and (c) for $x = 0$, 0.04 and 0.06, respectively.
For the superconducting composition $x = 0.06$, {\ibwlo} and {\ibwtr} in the superconducting state at $T = 3$~K do not differ much from those in the normal state at $T = 35$~K. 
This agrees with the earlier reports~\cite{Ino10, Par10}.
For nearly antiferromagnetic metals~\cite{Mor85}, temperature dependence of {\ibwlo} and {\ibwtr} is in proportion to $\Gamma(T)^{-1/2} \propto (T+\Theta)^{-1/2}$. 
The solid lines in figure~\ref{fig3} are the fitting results with $(T + \Theta)^{-1/2}$ with $\Theta = 30$~K. The function gives a good fit to {\ibwlo} and {\ibwtr} for $x = 0.04$ and 0.06, which is consistent with~\cite{Ino10}, however, a poor fit for $x = 0$. 
Increase of {\ibwlo} and {\ibwtr} for $x = 0$ is much rapid on decreasing temperature than those expected for nearly antiferromagnetic metals. 

\begin{figure}
\begin{center}
        \includegraphics[width=0.9\hsize]{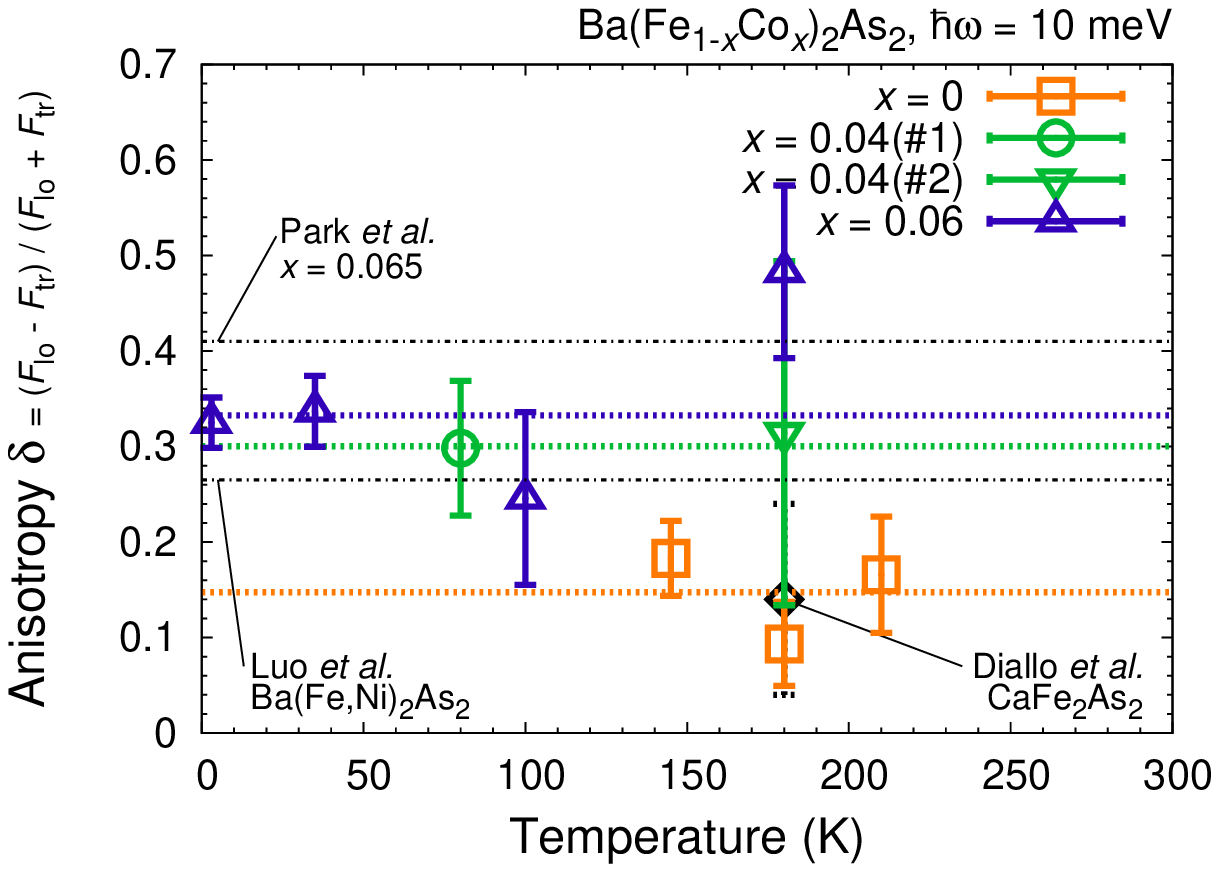}
        \caption{\label{fig4}(Colour online) Temperature dependence of the anisotropy of magnetic correlation $\delta$ around $\ibbs{Q} = (1/2, 1/2, 0)$ at ${\hw} = 10$~meV in {\ibbfcax}. The (orange) squares stand for $x = 0$, (green) circles for 0.04(\#1), (green) inverted triangle for 0.04(\#2), and (blue) triangles for 0.06. The (red, green and blue) thick dashed lines show the average values of the anisotropy $\delta_{\rm{ave}}$ for $x = 0$, 0.04 and 0.06(, respectively). The thin dash-dot lines are the anisotropy for $x = 0.065$~\cite{Par10}, and the average anisotropy for Ni-doped BaFe$_2$As$_2$~\cite{Luo12}. The diamond shows the anisotropy for CaFe$_2$As$_2$~\cite{Dia10}.} 
        \end{center}
\end{figure}
Next, we check the relation between {\ibwtr} and {\ibwlo} by defining the anisotropy ratio $\delta$ as $(\ibwlo - \ibwtr) / (\ibwlo + \ibwtr)$. 
Temperature dependence of $\delta$ is shown in figure~\ref{fig4} for $x = 0$, 0.04 and 0.06, respectively. 
$\delta$ is temperature independent in all the composition within the experimental uncertainty. 
The average anisotropy $\delta_{\rm{ave}}$ of each compound is 0.15(7), 0.30(13) and 0.33(6) for $x = 0$, 0.04 and 0.06, respectively. 
The $\delta_{\rm{ave}}$ of $x = 0$ is clearly smaller than those of electron-doped compounds $x = 0.04$ and 0.06.
$\delta_{\rm{ave}}$ of $x = 0$ is close to the anisotropy ${\sim} 0.14(10)$ in CaFe$_2$As$_2$ at $T = 180$~K and $\hw = 12 {\pm} 5$~meV, reported by Diallo {\it et al}~\cite{Dia10}.
$\delta_{\rm{ave}}$ of $x = 0.06$ is roughly consistent to the anisotropy 0.41(2) of $x = 0.065$ reported by Park {\it et al}~\cite{Par10}.
A slight increase of anisotropy by Ni doping including antiferromagnetic state is reported by Luo {\it et al}~\cite{Luo12}.

\subsection{Doping dependence of anisotropy at ${\hw} = 28$~{\rm meV}}
\begin{figure}
\begin{center}
	\includegraphics[width=0.9\hsize]{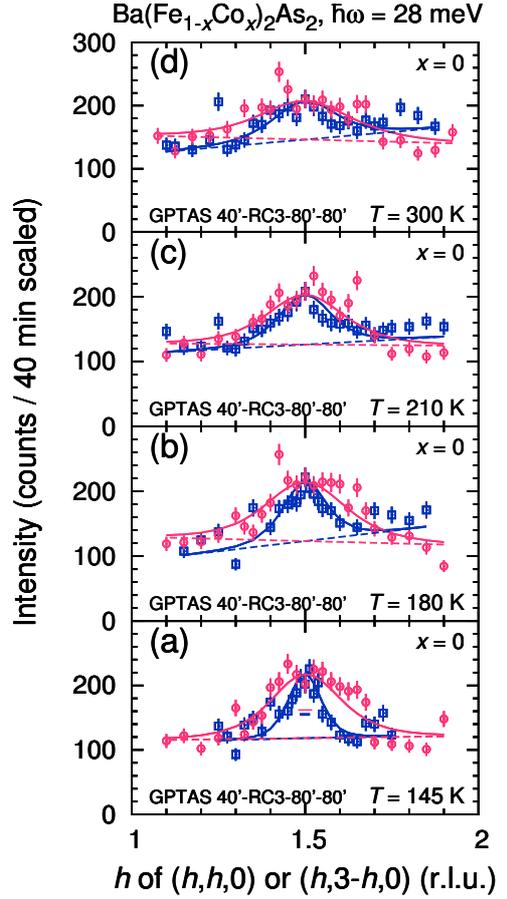}
	\caption{\label{fig5}(Colour online) (a)--(d) Longitudinal and transverse scans around $(3/2, 3/2, 0)$ in the paramagnetic state of BaFe$_2$As$_2$ at ${\hw} = 28$~meV at $T = 145$, 180, 210 and 300~K, respectively. The (blue) squares denote the longitudinal scan, and the (red) circles denote transverse scan. The solid lines indicate fits, and dashed lines show backgrounds in the fitting procedure. All data were measured with a counting time of 40 min or higher, but are normalized to counts/40 min.}
	\end{center}
\end{figure}

\begin{figure}
\begin{center}
\includegraphics[width=0.9\hsize]{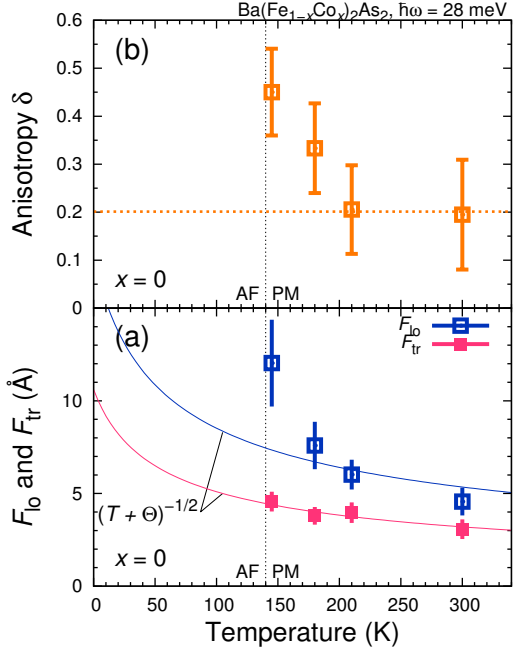}
\caption{\label{fig6}(Colour online) (a) Temperature dependence of {\ibwlo} and {\ibwtr} of $\ibbs{Q} = (3/2, 3/2, 0)$ at ${\hw} = 28$~meV in BaFe$_2$As$_2$. The solid lines indicate fit with $(T+\Theta)^{-1/2}$.
(b) Temperature dependence of the anisotropy $\delta$ around $\ibbs{Q} = (3/2, 3/2, 0)$ at ${\hw} = 28$~meV in BaFe$_2$As$_2$. The dashed line shows the average value of $\delta$ for $T \geq 210$ K. The vertical (black) dotted lines in (a) and (b) stand for $T_{\rm AF} = 140$~K.}
\end{center}
\end{figure}

At ${\hw} = 28$~meV, temperature dependence of the anisotropy was investigated around $\ibbs{Q} = (3/2, 3/2, 0)$ in $x = 0$. 
The results of the constant-energy scans at $T = 145$, 180, 210 and 300~K are shown in Figs.~\ref{fig5}(a)--(d), respectively.
It is clear in the figure that considerable $\ibbs{Q}$ dependence exists even at the room temperature $T = 300$~K $\sim 2T_{\rm{AF}}$. 
At all the temperatures, the transverse scans show larger peak widths compared to the longitudinal ones, in good agreement with the low energy results at $\hw = 10$~meV.
These data were fitted to (\ref{eq:chi}) with the fixed $D$ = 1.3~r.l.u.
In the fitting procedure, the backgrounds were assumed to be linear in the longitudinal and transverse directions, and resolution convolution was similarly performed as for the 10~meV data. 
Fitting results are shown by the solid lines in figure~\ref{fig5}.
Temperature dependence of {\ibwlo} and {\ibwtr} is shown in figure~\ref{fig6}(a). 
As inferred from the raw data in figure~\ref{fig5}, the obtained optimum parameter {\ibwtr} is smaller than the longitudinal {\ibwlo}. 
{\ibwlo} and {\ibwtr} gradually increase as the temperature is decreased. 
{\ibwtr} shows good fit with $(T+\Theta)^{-1/2}$, although {\ibwlo} shows poor fit as shown by the solid lines in figure~\ref{fig6}(a).
The ratio of anisotropy $\delta$ is shown in figure~\ref{fig6}(b).
$\delta$ is positive even at $T = 300$~K $\sim 2T_{\rm{AF}}$. $\delta$ at high temperatures are roughly the same as those at ${\hw} = 10$~meV. On the other hand, $\delta$ becomes large at $T = 145$~K just above $T_{\rm{AF}}$; this behaviour is different from that at ${\hw} = 10$~meV.

\section{Discussion}
Three key findings in the present study are summarized as follows.
First, anisotropy of the inplane spin correlation $\delta$ is increased by electron doping from $x = 0$ to $x = 0.06$.
Secondly, for $x = 0$, the inplane anisotropy $\delta$ at ${\hw} = 28$~meV becomes large at $T = 145$~K, although $\delta$ above $T = 210$~K is as small as that at ${\hw} = 10$~meV. 
Thirdly, for $x = 0$, {\ibwlo} and {\ibwtr} increase more than those expected for nearly antiferromagnetic metals as temperature decreasing, with the inplane anisotropy keeping constant within errors at ${\hw} = 10$~meV. 

First and second results are consistent with a simple RPA calculation taking the multiorbital character of Fe $3d$ bands into account~\cite{Gra10, Par10}.
The increasing behavior of the anisotropy with the electron doping is indeed expected in the earlier study~\cite{Par10}.
The larger inplane anisotropy at $T = 145$~K compared to that at 210~K for the energy ${\hw} = 28$~meV may be understood as follows:
the nesting condition may be more anisotropic at higher energy transfer, as inferred in a high-energy study by Harriger {\it et al}~\cite{Harr11}. 
Such anisotropic nesting may be clearly seen in the spin excitations at lower temperature, however, at high temperatures, thermal fluctuations may smear the details of electronic structure around the Fermi level.
This would be the reason why the observed the pronounced anisotropy at lower temperatures, whereas spectra become similar to those at ${\hw} = 10$~meV.
Hence, both the lower- and higher-energy results indicate that the anisotropic antiferromagnetic fluctuations are mostly dominated by the Fermi surface nesting.

In contrast to first and second results, the temperature dependence of the peak width for the parent compound is inconsistent with that expected for nearly antiferromagnetic metals.
A similar discrepancy can be found in another parent compound CaFe$_2$As$_2$ reported by Diallo {\it et al}~\cite{Dia10}. 
It should be noted that the peak sharpening is naturally expected as the  temperature becomes closer to the transition temperature. What is truly unusual here is that the temperature dependence of the peak width for $x = 0$ breaks the relation to the damping constant, {\it i.e.}, $F \propto \Gamma^{-1/2}$. 
The previous neutron inelastic scattering study reported~\cite{Mat10} that the critical slowing down process of the spin fluctuations, $\Gamma(T)$, is interrupted at $T_{\rm{AF}}$, implying the first-order magnetic phase transition. The interruption is also observed by the NMR technique through the spin-lattice relaxation rate~\cite{Kita08, Ning09}. 
For small $\ibbs{q}$ and $\hw$ region of itinerant antiferromagnets~\cite{Mor85}, the peak width $F$ is connected to the damping constant $\Gamma(T)$ with $\Gamma(T) \propto F^{-2}$. 
Therefore, in the absence of a large critical slowing down, the peak width is expected to show relatively moderate temperature dependence even immediately above the transition temperature.
This disagreement may be another appearance of the unconventional critical behaviour of the first-order magnetic transition in the parent compounds. 
For $x \le 0.022$, the high resolution X-ray~\cite{Kim11} and magnetization~\cite{Rotu11} measurements demonstrated that the first-order combined magnetic and structural transition follows the second-order structural transition on decreasing temperature, suggesting the existence of the magnetic tricritical point~\cite{Cano10} at $x \sim 0.022$. Neutron diffraction measurements showed that the antiferromagnetic order parameter exponent $\beta$ is different between below and above the tricritical point; $\beta$ is about the two-dimensional Ising value of 0.125 for $x = 0$~\cite{Wil09, Wil10}, the mean-field tricritical value of 0.25 for $x = 0.021$ and 0.022~\cite{Paje13}, and three-dimensional Ising value of 0.327~\cite{Camp02} for under-doped compounds for 0.039~\cite{Paje13} and $x = 0.047$~\cite{Pra09}.
The inconsistency of the peak width may represent the existence of unconventional spin dynamics even below the tricritical point $x \le 0.022$. 

Supplementary, we note that the dynamical susceptibility $\chi''$ calculation, on which our above arguments are based on, may have quantitative shortfalls.
Indeed, the inplane peak in the calculated $\chi''$~\cite{Gra10, Par10} is significantly broader than the observed one.
The orbital distribution of the Fe $3d$ bands used in the above RPA calculation is not consistent with the recent ARPES result~\cite{Zha11}, and hence the quantitative validity of the RPA calculation may be questioned also from this viewpoint.
Nonetheless, the doping dependence of the anisotropy in the nesting picture is caused by the different sizes of the hole and electron pockets, and hence, smaller anisotropy in the parent compound only requires a smaller difference of the pocket sizes. 
Therefore, a minor modification of the calculated Fermi surface will not affect the conclusion.

\section{Summary}
We investigated antiferromagnetic spin fluctuations in {\ibbfcax} crystals ($x = 0$, 0.04 and 0.06) by inelastic neutron scattering technique. The inplane anisotropy was clearly observed for the low-energy spin fluctuations. The anisotropy is larger in the higher doping level. 
The result agrees with the RPA calculations including the orbital character of the electron bands~\cite{Par10, Gra10}. 
The large anisotropy observed for ${\hw} = 28$~meV would also be explained within the Fermi surface nesting picture. 
We conclude that the inplane anisotropy of the spin correlations is mostly dominated by the Fermi surface nesting. 
Concerning the temperature dependence of the peak width, the doped compounds show consistent behavior expected for nearly antiferromagnetic metals, whereas the peak in the parent compound sharpens much pronouncedly.
This suggests that the quasi-two-dimensional spin correlations grow much rapidly for decreasing temperature in the $x = 0$ parent compound.
This may be another sign of the unconventional nature of the antiferromagnetic transition in BaFe$_2$As$_2$.
For quantitative discussion of the anisotropy, further study of the electronic structure near the Fermi level is necessary.

\section{Acknowledgments}
The authors thank M. Rahn for supporting our neutron scattering experiments and M. Imai for helpful comments. 
This work is partly supported by the U.S.-Japan cooperative program on neutron-scattering research. Part of this work was supported by the Division of Scientific User Facilities, Office of Basic Energy Sciences, U.S. DOE, and KAKENHI (23340097 and 23244068) from MEXT, Japan.

\bibliographystyle{model1-num-names}
\bibliography{ibuka20131017-notitle}

\end{document}